\documentclass[a4paper]{article}
\usepackage{algorithmic}
\usepackage{textcomp}
\usepackage{xcolor}
\def\BibTeX{{\rm B\kern-.05em{\sc i\kern-.025em b}\kern-.08em
    T\kern-.1667em\lower.7ex\hbox{E}\kern-.125emX}}

\usepackage{balance}
\usepackage{hyperref}
\usepackage{natbib}
\usepackage{bbm}

\def\BibTeX{{\rm B\kern-.05em{\sc i\kern-.025em b}\kern-.08em
    T\kern-.1667em\lower.7ex\hbox{E}\kern-.125emX}}
\usepackage{amsmath}
\usepackage{amssymb}
\usepackage{amsfonts}

\usepackage[colorinlistoftodos]{todonotes}

\usepackage{xspace}

\usepackage{subcaption}
\newcommand\etal[0]{\emph{et al.}\xspace}

\usepackage{framed}
\definecolor{shadecolor}{rgb}{1,0.8,0.3}

\usepackage{algorithm}
\usepackage{algorithmic}

\usepackage{graphics}

\begin{document}

% \title{Roadmap to Uniformity: Testing Uniform Random Samplers}
\title{Testing Uniform Random Samplers: Methods, Datasets and Protocols}

\author{Olivier Zeyen}
% \affiliation{%
%   \institution{University of Luxembourg, SnT}
%   \country{Luxembourg}
% }

\author{Maxime Cordy}
% \affiliation{%
%   \institution{University of Luxembourg, SnT}
%   \country{Luxembourg}
% }

\author{Martin Gubri}
% \affiliation{%
%   \institution{Parameter Lab}
%   \country{Germany}
% }

\author{Gilles Perrouin}
% \affiliation{%
%   \institution{PReCISE/NaDI, University of Namur}
%   \country{Belgium}
% }

\author{Mathieu Acher}
% \affiliation{%
%   \institution{Univ Rennes, Inria, CNRS, IRISA}
%   \country{France}
% }

\author{Olivier Zeyen, University of Luxembourg, SnT\\
Maxime Cordy, University of Luxembourg, SnT\\
Martin Gubri, Parameter Lab\\
Gilles Perrouin, PReCISE/NaDI, University of Namur\\
Mathieu Acher, Univ Rennes, Inria, CNRS, IRISA
}

\maketitle

\begin{abstract}
Boolean formulae compactly encode huge, constrained search spaces.
Thus, variability-intensive systems are often encoded with Boolean formulae.
The search space of a variability-intensive system is usually too large
to explore without statistical inference (e.g. testing).
Testing every valid configuration is computationally expensive (if not impossible)
for most systems. This leads most testing approaches to sample a few configurations
before analyzing them.
A desirable property of such samples is \textit{uniformity}:
Each solution should have the same selection probability.
Uniformity is the property that facilitates statistical inference.
This property motivated the design of uniform random samplers, relying on SAT
solvers and counters and achieving different trade-offs between uniformity and scalability.
Though we can observe their performance in practice, judging the quality of the
generated samples is different. Assessing the uniformity of a sampler
is similar in nature to assessing the uniformity of a pseudo-random number (PRNG)
generator. However, sampling is much slower and the nature of sampling
also implies that the hyperspace containing the samples is constrained.
This means that testing PRNGs is subject to fewer constraints
than testing samplers.
We propose a framework that contains five statistical tests which
are suited to test uniform random samplers.
Moreover, we demonstrate their use by testing seven samplers.
Finally, we demonstrate the influence of the Boolean formula
given as input to the samplers under test on the test results.

\end{abstract}

\section{Introduction}
\label{sec:intro}
Uniform Random Sampling (URS) is the problem of generating
random SAT-solutions from a Boolean formula, such that every solution of the input formula gets a
uniform probability of being returned. 
URS has many applications in software engineering
including software product lines (SPL) and software testing.
SPLs often have huge configuration spaces due to combinatorial explosion. 
Thus to test an SPL, testing every configuration is often
intractable even for small numbers of features.
For example, JHipster has 45 features which result in 26256 possible configurations.
Testing every configuration would be very time-consuming or even
unfeasible depending on the available computation budget.
One may thus use URS to obtain an unbiased sample
of the configuration space and only test these configurations.
While this may not be exhaustive,
the sample size can be adjusted to account for the computation
budget and the desired confidence in the test results
\citep{Oh2017,DBLP:conf/splc/OhGB19,Plazar2019}. 
URS may also be used to search for optimal configurations
as shown in \citep{Oh2017}.
Other applications include the use of URS to learn the configuration
space \citep{pereira2021learning} where URS is used to generate
an unbiased dataset which is then used to train a model.
Deep learning verification \citep{DBLP:conf/icse/BalutaCMS21}
and evolutionary algorithms \citep{dePerthuisdeLaillevault:2015:MNS:2739480.2754760}
have also been shown to benefit from URS.

The key role of URS in these applications is that uniformity enables unbiased sampling from a large solution space. 
Some samplers provide theoretical guarantees of uniformity
thereby proving that their algorithm generates samples
with uniform probability \citep{Achlioptas2018FastSO,SGM20}. However, these proofs of uniformity are hard to provide for multiple reasons.
First, some samplers favor practical heuristics that trade-off potential theoretical guarantees for improved efficiency \citep{STS}.
% Second, the theoretical guarantees may be impossible to obtain
% regardless of the uniformity of the sampler
% due to Gödel's incompleteness theorems \citep{godel1931formal}.
% In other words, there may exist a sampler $S$
% which is theoretically uniform but for which the theoretical
% proof of uniformity cannot be made.
Second, the theoretical guarantees are done on the algorithm
and not on the implementation of an algorithm, which may break the theoretical guarantees due to the way the sampler is implemented, or merely if the implementation contains undetected bugs.

In this paper, we seek to establish a methodological ground for testing sampler uniformity in practice, by means of a set of statistical tests. Our contributions not only enable researchers to compare practical uniformity confidence across multiple sampler candidates, but they can also be useful to detect implementation mistakes leading to non-uniform tools although based on a theoretically uniform method. The statistical tests composing our method constitute an actionable solution for URS researchers to assess the uniformity of novel approaches while removing the necessity of developing arbitrarily complex (and, as argued before, only partially reliable) theoretical proofs. All in all, our contribution aims to foster and accelerate research in URS by automating empirical testing and enabling comparison with a wide range of
existing samplers.

To achieve this, we inspire ourselves from the
pseudo-random number generator (PRNG) community \citep{l2007testu01,PractRand,brown2018dieharder}, which has solid experience in practical testing approaches.
Unfortunately, we cannot reuse these PRNG tests because the solution space of PRNG is unconstrained, unlike Boolean formulae in URS. Nevertheless, we follow the good practices of the PRNG community to establish our own test protocol. In particular,
this community has long used multiple statistical
tests to test their tools, as each test has different strengths
and weaknesses \citep{rukhin2001statistical}.
Relying on a single test $A$ is bad practice,
as a PRNG could be engineered to specifically pass test $A$
without necessarily producing high-quality
pseudo-random numbers.

Applying this lesson learned from the PRNG community, we start by providing a statistical test suite
which is designed to test the uniformity of uniform random samplers. These tests have been selected and adapted from the available literature to form a consistent testing approach for URS. After describing the different tests that make up our suites, we apply them in an empirical study involving the available URS tools. Throughmar chthis study, we can not only compare the uniformity test results for the samplers but also reveal the consistency of all tests (or lack thereof) on a given sampler. Through these investigations, we confirm the existence of an experimental bias in the choice of a specific test and, thereby, the necessity of conducting different tests. 

Through our applications of our tests to state-of-the-art URS tools, we find that different tests not only have a different reliability but also come with varying computational cost. Based on these insights, we suggest a testing process that executes different tests in a sequence such that tests with a lesser computational cost -- and a potentially higher Type-II error rate\footnote{Our null hypothesis is that a given sampler is uniform. Therefore, Type-I error refers to the rejection of this hypothesis while the sampler is uniform. Type-II error refers to the conclusion that a sampler is uniform while it is not.} -- are executed first. We also indicate which of these tests reveal less information regarding sampler uniformity and can be omitted in the case of constrained computational resources.

Next, we explore another threat to the validity of uniformity testing: the input Boolean formula used for testing. Indeed, while one could consider uniformity as a universal property of sampling methods, the empirical nature of both sampler implementation and uniformity tests creates an inherent risk for a test to generate Type-I and Type-II errors. We, therefore, consider the use of multiple formulae for uniformity testing and present a methodology to combine results for individual tests into a statistically meaningful answer. Beyond this, we study the question of dataset bias.
In particular, we investigate how test results obtained from synthetic formulae
typically used in the URS research community
correlate with results on real-world formulae extracted from SPL feature models. Establishing these correlations (or their absence) would determine the importance and limitations of synthetic datasets in uniformity testing.

% one can look at the uniformity of a sampler as a universal
% property. Meaning that if a sampler is uniform then it is uniform regardless
% of the input formula. Thus the question rises: Is the input formula important?
% A negative answer would mean that we can obtain reliable results
% even when using small (or even synthetic) formulae which would drastically
% cut the computational cost of statistical testing.
% A positive answer however would mean that sampler testing needs to be
% performed on a wide range of different formulae to obtain reliable results.
% Thus the used dataset should be chosen with care.

To summarize, this paper makes the following contributions: 
\begin{enumerate}
    \item \textbf{A uniformity testing procedure.} Our method includes five tests, of which three have never been applied as statistical tests for sampling uniformity. Our procedure also considers the relative computational costs of the tests in order to quickly eliminate non-uniform samplers and limit the execution of more expensive tests.
   
    \item \textbf{Uniformity results on state-of-the-art samplers.}
    We report on a large experimental study studying the uniformity of state-of-the-art samplers according to our different tests.
    We reveal that among our tested samplers, only UniGen3 is uniform,
    unlike what previous studies concluded. These results show the importance of conducting multiple independent tests to ensure reliability, while also avoiding the pitfall of previous approaches in using a supposedly uniform sampler as a reference \citep{SGCM22}.
    
    % \zyno{This is important as Barbarik uses a reference sampler.
    % Our uniformity results demonstrate the importance of our tests
    % compared to Barbarik.}
    % This further highlights the importance of uniformity testing and software
    % testing in general.
    % Moreover, our results show that computationally expensive tests
    % are not necessarily more reliable than more expensive tests thus
    % showing that some tests may be executed only if previous (cheaper) tests
    % are validated first to allow for faster and cheaper testing.
    \item \textbf{Insights into dataset choice.}
    We reveal that URS tools that fail to sample uniformly from synthetic formulae are also unlikely to produce uniform samples from real-world formulae. The contraposition, however, does not hold, revealing the need to benchmark samplers against a diverse set of real-world formulae.
    
\end{enumerate}

% \textbf{{Open science policy.}} All our experimentation infrastructure is available at the following link: \url{}.

\section{Related Work}
\label{sec:related}
% \paragraph*{From Uniform Sampling to Uniformity Testing and Back}

As indicated by \citet{Plazar2019}, assessing the uniformity
distribution of SAT solutions is difficult and a direct method is prohibitively
expensive.
Barbarik by \citet{SGM20} is a test which, by using
a uniform random sampler as reference tests whether a given input sampler is uniform or not.
In other words, Barbarik tests whether both samplers sample from the
same distribution. If the reference sampler is uniform, then Barbarik
tests the uniformity of the sampler under test.
The authors updated Barbarik \citep{SGCM22} to support a more 
fine-grained analysis of uniformity.
The approach has the main downside of requiring a uniform random sampler
as a reference. If the reference sampler is not uniform then the results
are unreliable. A high level of trust is thus required for the
reference sampler.
Another way of assessing the solutions' uniformity is the statistical test
proposed by \citet{BDDSampler,DBLP:conf/splc/HeradioFGB20}.
% \citep{DBLP:conf/splc/HeradioFGB20,BDDSampler}.
While 
the two kinds of approaches seem to agree on the (non-)uniformity of most samplers, 
they seem to disagree on Smarch's status \citep{SMARCH}.
It is currently an open question of whether 
this disagreement stems from a different test design or the selection of SAT formulae 
which, though overlapping, are not exactly the same. We note that there is a close 
relationship between the design of URS techniques and testing uniformity: the 
improvements of Barbarik led to CMSGen, while Heradio \etal's uniformity tests 
led them to develop BDDSampler, a novel uniform sampler based on binary decision 
diagrams \citep{BDDSampler}. We excluded the latter from this study due to 
its SAT-based complexity focus.

Assessing the quality of a pseudo-random number generator (PRNG)
is similar to assessing the quality of a uniform random sampler.
One key property desired by both is that every possible value
should have a uniform probability of being returned.
In the case of URS, this means returning a solution to a Boolean
formula at random. In the case of PRNGs, this means returning
an integer (on a usually predefined number of bits)
uniformly at random. In other words, a PRNG is a uniform random sampler
for a Boolean formula with no constraints.
Statistical testing of PRNGs to assess their quality
is quite common. One test suite is the NIST test suite
developed by \citet{rukhin2001statistical}.
Other, more recent test suits include TestU01 \citep{l2007testu01},
dieharder \citep{brown2018dieharder} and PractRand \citep{PractRand}.
All of these test suites contain multiple statistical tests.
Each test has different strengths and weaknesses.
Thus each test is suitable to detect different kinds of
weaknesses in the generated stream of numbers.
For example, the monobit test checks whether the number
of ones and zeroes is approximately equal in the binary
stream of numbers generated by a PRNG.
If a PRNG generates a stream of alternating ones and zeroes
then it would pass the monobit test. However, the PRNG
would produce "random" numbers of poor quality thus
highlighting the need for multiple tests.

Unfortunately, most of the tests performed on PRNGs
rely on the fact
that the solution hyperspace is unconstrained.
Moreover, PRNGs are much faster at generating samples
than uniform random samplers.
This is shown by \citet{blackman2021scrambled}
who tested PRNGs on up to $10^{15}$
generated bytes. This is feasible as PRNGs are engineered
to generate 32-bit or 64-bit words in the nanosecond
or even sub-nanosecond range. However,
URS implementations often rely on NP-Oracles
such as SAT-solvers \citep{STS} or on \#P-Oracles
\citep{Achlioptas2018FastSO,SGRM18,valiant1979complexity}.
This implies that URS is much slower than pseudo-random
number generation.
Thus, URS needs its own set of specialized tests
that can work with small sample sizes
and under the constrained hyperspace implied
by the Boolean formula given as input to the sampler
under test.

We contribute a set of five tests, some are already
in the URS literature and have been adapted to our use case
and others needed to be adapted to URS.
We are thus, to the best of our knowledge, the first
to use multiple tests on uniform random samplers.

We are also the first to explore
how the Boolean formulae provided as input influence the test outcomes.
PRNG testing focused
on statistical testing, but uniform random samplers
expect a Boolean formula as input.
Therefore, it is important to study how the input affects
the results and any potential noise that may arise in the test results.

% \citet{l2007testu01}
% \citet{brown2018dieharder}
% \citet{ONeillbday}
% \citet{PractRand}
% \citet{blackman2021scrambled}

\section{Background}
\label{sec:background}
\subsection{Boolean formulae}

A Boolean formula $F$ is defined over a set of Boolean variables $\textit{Var}(F)$ and takes a Boolean value that can evaluate to either true (1) or false (0). A literal is either a variable $x \in \textit{Var}(F)$ or its negation $\neg x$, such that if variable $x$ is set to true then the literal $x$ evaluates to true and the literal $\neg x$ evaluates to false.
We use the notations $\textit{Var}(x)$ and $\textit{Var}(\neg x)$ to refer to the variable corresponding to the literal $x$ and $\neg x$, respectively.

An assignment $m$ to $Var(F)$ is called a model of $F$ if and only if $m \models F$.
We denote the value of the variable $v$ in $m$ by $m(v)$.
We denote the set of models of $F$ by $R_F$.
Thus $m \models F$ is equivalent to $m \in R_F$.
We define $|R_F|$ as the size of the set $R_F$.\\

A formula $F$ is in negational normal form (NNF) if
negation is used only on literals. Furthermore,
it is in conjunctive normal form (CNF) if written as a conjunction
of disjunction of literals ($F = \bigwedge_{A_i} \bigvee_{l \in A_i} l$).
A d-DNNF is an NNF form where every conjunction is
\emph{decomposable} and every disjunction is \emph{deterministic} \citep{D4}.
A conjunction $\bigwedge N_i$ is decomposable if
for every pair $(j,k)$ we have $j \neq k \implies \textit{Var}(N_j) \cap \textit{Var}(N_k) = \emptyset$. A disjunction $\bigvee N_i$ is deterministic if for
every pair $(j,k)$, $j \neq k \implies R_{(N_j \land N_k)} = \emptyset$.

We next define three common problems defined over Boolean formulae, i.e. SAT solving, model counting, and URS.
SAT solving is the problem of finding a model for a given
formula $F$ such that $m \models F$.
Model counting (\#SAT) is the problem of
computing $|R_F|$.
Uniform random sampling (URS) is the problem of
sampling a model from $R_F$ such that every $m \in R_F$ has a probability
$\frac{1}{|R_F|}$ of being sampled.

Despite their intrinsic links, these three problems are different and require dedicated solutions to be addressed. For example, any sampler can be used for SAT solving because they have to generate at least one model of the formula; however, SAT solvers offer no uniformity guarantee because they are only interested in proving the existence of one model.

%\gpe{Miss a paragraph or two about samplers and their relevance for the SE/AI/FM fields}

\subsection{Samplers}

% \zyno{Unsure if we should keep the distaware sampler, the testing process
% may be flawed regarding this sampler as it does not seem to take
% a random seed as parameter, this could explain why it returns so many
% repetitions in the birthday test.}
% \textbf{Distance-based Sampling} \citep{Kaltenecker2019}: an algorithm
% using a distance metric to guide an off-the-shelf SAT solver towards different
% areas of the solution space. The distance metric used is the number of
% variables set to true. This generates additional constraints
% which are added to the input formula and the SAT solver generates a solution.

\textbf{KUS} \citep{SGRM18}: a sampler for d-DNNF. d-DNNF is a special kind of negational
normal form for propositional formulae which allows for polynomial time model enumeration
and model counting, both of which are interesting for sampling. Thus, KUS uses recent advances
in knowledge compilation to compile a formula to d-DNNF and then samples from the compiled
formula.
% This makes it a uniform sampler by design.

\textbf{QuickSampler} \citep{Dutra2018}: an algorithm based on a strong set of heuristics, which are shown to produce samples quickly in practice on a large set of industrial benchmarks \citep{Dutra2018}. However, the tool offers no guarantee on the distribution of generated samples, or even on the termination of the sampling process and the validity of generated samples (they have to be checked with a SAT solver after the generation phase).

% oz
\textbf{Smarch} \citep{SMARCH}: Smarch recursively partitions the configuration
space using a model counter and a random number generator. At each step, Smarch
counts the total number of solutions of a formula. Smarch
then picks a random integer smaller than the number of solutions,
and depending on its value and the number of solutions
with a given variable set to false, it either sets the variable to true or false.
%Thus Smarch is uniform by design.

% oz
\textbf{SPUR} \citep{Achlioptas2018FastSO}: SPUR is built on top of sharpSAT \citep{sharpSAT},
a model counter.
SPUR takes advantage of the fact that sharpSAT systematically
explores all the solutions of a formula to sample from it.
As SPUR is tightly integrated with sharpSAT, it can leverage this
process to generate uniform samples. Furthermore, SPUR is one of
the few samplers that provides theoretical guarantees regarding its uniformity.

% oz
\textbf{STS} \citep{STS}: a SAT solver based algorithm.
% While sampling using a DPLL algorithm is not uniform, STS has guarantees on the uniformity of the samples.
STS samples partial solutions recursively.
At each step, the partial solutions are extended, and only
the valid partial solutions are kept.
If the number of partial solutions exceeds a threshold,
a sub-sample is randomly selected.
STS continues
until all variables have been assigned, which results in a set of solutions
of the initial formula. STS uses a SAT solver to verify the validity
of the partial assignments.

\textbf{CMSGen} \citep{CMSGen}: The authors used a uniformity testing tool
called Barbarik \citep{Chakraborty2019OnTO} to adjust the parameters of the SAT solver
CryptoMiniSAT \citep{Soos2009ExtendingSS} until it showed satisfying uniformity.

% \textbf{UniGen} \citep{chakraborty2013scalable,Chakraborty2015ug2}: a hashing-based algorithm to generate samples in a nearly uniform manner with strong theoretical guarantees: it either produces samples satisfying a nearly uniform distribution or it produces no sample at all. These strong theoretical properties come at a cost: the hashing based approach requires adding large clauses to formulae so they can be sampled. These clauses grow quadratically in size with the number of variables in the formula, which can raise scalability issues.
% 
% % oz
% \textbf{UniGen2} \citep{Chakraborty2015ug2}: a hashing-based algorithm based on UniGen. UniGen2 weakens the guarantee of independence among samples to achieve faster sampling when compared to UniGen. This also allows UniGen2 to be parallelized with a near-linear speedup in the number of cores used.

% oz
\textbf{UniGen3} \citep{SGM20}: a hashing-based algorithm to generate samples in a nearly uniform manner with strong theoretical guarantees: it either produces samples satisfying a nearly uniform distribution or it produces no sample at all. These strong theoretical properties come at a cost: the hashing-based approach requires adding large clauses to formulae so they can be sampled. These clauses grow quadratically in size with the number of variables in the formula, which can raise scalability issues.
% a hashing-based algorithm. To improve on UniGen2 the authors
% investigated the bottlenecks of UniGen2 and made key improvements to their algorithm
% and to the way CryptoMiniSat handles XOR frormulae which led to better performance.

BDDSampler \citep{BDDSampler} is a more recent sampler that
operates on binary decision diagrams. As such it is theoretically
uniform. We exclude BDDSampler from our study because it is
similar to KUS. However, unlike KUS which takes a CNF formula in DIMACS
format, the BDDSampler implementation expects a binary decision
diagram as input.

\section{Statistical Test Methodology}
\label{sec:approach}
Our objective is to provide a statistically grounded method to test the uniformity of URS tools. The key assumption of our method is that uniformity testing typically suffers from two pitfalls: the reliance on a single, specific test and the biases introduced by the choice of the input formula. Given these pitfalls, our approach advocates the use of several statistical tests and the combination of results obtained from different formulae into a single statistical answer.

\subsection{Statistical Tests for Uniform Random Sampling}

We propose five statistical tests to empirically assess sampler uniformity. The intuition behind the use of multiple tests is that each test assesses whether a particular statistical value from the solution space is preserved in the sample. Therefore, the result for a single test is bound to the statistical value it evaluates, which makes it necessary to execute multiple tests to obtain higher confidence regarding sampling uniformity. To build the tests, we consider the following hypotheses: 
\begin{itemize}
    \item $H_0$ (null hypothesis): the sampler samples uniformly,
    \item $H_a$ (alternative hypothesis): the sampler does not sample uniformly.
\end{itemize}

\subsubsection{Pearson's $\chi^2$ Statistical Test (GOF)}

The process of examining how well a sample agrees with a probability distribution
is known as a \emph{goodness of fit} test. We describe
below how one can apply Pearson's
$\chi^2$ test \citep{Pearson1900OnTC} to uniform random samplers.

Person's $\chi^2$ test requires two vectors $E$ and $O$.
The vector $E$ contains the expected data under $H_0$
and $O$ contains the observed data of a sample $S \subseteq R_F$ of size $|S| = N$.

We define $O = (O_1, O_2, ..., O_{|R_F|})$ for a sample $S$
as $O_i = |\{m \in S | id(m) = i\}|$. With $id$, a bijective function
$id : R_F \rightarrow \{1, 2, ..., |R_F|\} \subseteq \mathbb{N}$.
We follow by defining $E = (E_1, E_2, ..., E_{|R_F|})$
as $E_i = \dfrac{N}{|R_F|}$. This follows from the definition of $H_0$.
As we want uniform probability, we expect the same number of occurrences
for every solution.

Person's $\chi^2$ test statistic is defined as follows:
\begin{equation}
\chi^2 = \sum_{i=1}^{|R_F|} \dfrac{(O_i - E_i)^2}{E_i}
\end{equation}

The p-value for Pearson's one-sided $\chi^2$ test
(with $|R_F| - 1$ degrees of freedom)
can be derived from
the statistic.
The p-value is then compared to a desired significance level $\alpha$.
If the p-value is smaller than $\alpha$, $H_0$ is rejected and $H_a$ is accepted.
Otherwise, no clear conclusion can be reached and we fail to reject $H_0$.
In our case, we will simplify the conclusion by accepting $H_0$.

% \zyno{Find source:
% The approximation to the $\chi^2$ distribution is known to break down if any expected
% frequency is below 5 ($E_i < 5$). \citep{Yates1934ContingencyTI}}

% This adaptation of Pearson's $\chi^2$ test will perform as expected.
% However, the test requires a large number of samples (at least 5 times the number
% of solutions because of the above reason). This large number of samples
% is intractable for larger formulae with an already high number of solutions.
% Moreover, one has to keep an array containing the observed frequencies of each
% solution in memory during sampling which may be impossible if $|R_F|$ is large.
% 
The main disadvantage of the test is that it works
with the entire probability distribution.
Thus it requires an array of elements representing the probability
distribution. This means one value for each possible output.
In the case of a PRNG, this means having an array of size $2^{32}$
or $2^{64}$ in the case of 32-bit or 64-bit PRNGs respectively.
In the case of Boolean formulae, the problem is similar,
the hyperspace is often too large which prohibits us from
allocating an array of the required size
as a lot of formulae have more than $2^{64}$ solutions.
Moreover,  the GOF test is known to be unreliable if any expected
frequency is below 5 \citep{Yates1934ContingencyTI}.
In other words, in the case of uniform random sampling,
the test requires us to sample at least five times the total number of solutions
to get reliable results.
Thus, GOF only scales to small hyperspaces.

\subsubsection{Monobit Statistical Test}

Next, we detail our adaptation of
the PRNG monobit statistical test \citep{Rukhin2000AST}.
The simplicity of the test allows us to easily adapt it to URS.
The focus of the test is to
compare the number of ones and zeroes in a series of bits generated
by a pseudo-random number generator (PRNG).
Assuming the sequence of bits was generated randomly,
we would expect the average number of ones and zeroes to be roughly equal.
The test thus assesses how likely the observed number of
ones and zeroes is.

This statistical test assumes that the number of ones
and zeroes are roughly equal under $H_0$.
This assumption may not hold as we are working with constrained
hyperspaces. Therefore the test has to be adapted
to work on constrained hyperspaces such as the solution space
of a Boolean formula.
In the case of constrained hyperspaces, using a uniform random sampler
as a random bit generator is not appropriate, as there is no guarantee
that the constrained hyperspace would allow for approximately the same
number of ones and zeroes.

Our strategy for this test is the following.
We define the number $c_{\textit{even}}$ of solutions
in $R_F$ that have an even number of variables set to true as follows:

\begin{equation}
c_{\textit{even}} = |\{m \in R_F | \; 1 = \mathbbm{1}_{\textit{even}}(m) \}|
\end{equation}

\begin{equation}
\mathbbm{1}_{\textit{even}}(m) = 
\begin{cases} 
  1 & \textit{if} \quad |\{v \in Var(F) | \; m(v) = \textit{true} \}| \in 2\mathbb{N} \\
  0 & \textit{otherwise}
\end{cases}
\end{equation}

With $2\mathbb{N} = \{2x | \; x \in \mathbb{N}\}$ and $\mathbb{N}$ the
set of integers.
We define $c_{\textit{uneven}} = |R_F| - c_{\textit{even}}$,
the number of solutions with an uneven number of variables set to
true.

To test our sampler we sample $N$ solutions.
Let $S$ be the multiset containing the sampled solutions such that $|S| = N$.
We compute
$O_{\textit{even}} = |\{m \in S | \; 1 = \mathbbm{1}_{\textit{{even}}(m) } \}|$
as the number of observed solutions that have an even number
of selected variables and
$O_{\textit{uneven}} = N - O_{\textit{even}}$
as the number of observed solutions
that have an uneven number of selected variables.
We follow by defining $E_{\textit{even}} = N \dfrac{c_{\textit{even}}}{|R_F|}$
and $E_{\textit{uneven}} = N \dfrac{c_{\textit{uneven}}}{|R_F|}$.

We compute Person's $\chi^2$ test statistic as follows:
\begin{equation}
\chi^2 = \sum_{i \in \{\textit{even}, \textit{uneven}\}} \dfrac{(O_i - E_i)^2}{E_i}
\end{equation}

A p-value $p$ is derived by performing Pearson's one-sided $\chi^2$ test
(with one degree of freedom).
We reject $H_0$ if we find $p \leq \alpha$
with $\alpha$, a predefined significance level.

The main strength of this test is its scalability.
Since there are only two categories (even and uneven),
we can expect the required sample size $N$ for the test to be relatively small.
(unless $c_{\textit{even}} \ll c_{\textit{uneven}}$
or $c_{\textit{even}} \gg c_{\textit{uneven}}$).
However, the reliability of the test is also limited by the fact that there
are only two categories. The test is unable to detect
the non-uniformity of a sampler if the sampler
under test somehow generates a sample with the right
$O_{\textit{even}}$ and $O_{\textit{uneven}}$ values. This may happen
if a sampler was specifically designed to have this
property.
To truly assess the uniformity of a sampler, other more reliable tests
are necessary.

\subsubsection{Variable Frequency (VF) Statistical Test}
\label{subs:vfst}

\citet{Plazar2019}
developed a test in which they compare the expected variable
frequencies with the observed variable frequencies of the sample.
We extend their work
by replacing the fixed threshold with a GOF test.
This allows us to have more reliable results.

A uniform sample is expected to be representative
of the set of solutions. Thus, the observed variable frequencies should
not deviate significantly from the expected frequencies.
The variable frequency test is a variation of our
monobit test applied to each variable individually.
We simply substituted the function $\mathbbm{1}_{\textit{even}}(m)$
with $\mathbbm{1}_v(m) = m(v)$.
The VF test only tests that the observed variable frequencies
are representative of the real variable frequencies.
While this is not enough to certify true uniformity, it may be enough depending
on the use case. As an example, suppose that a variable $a$ is of interest
because the software component $a$ has a bug. If we know the variable
frequency of $a$, then we have an idea of how widespread the problem
is relative to the entire software product line.

Next, we show how to modify the monobit test
to obtain a variable frequency test.
The resulting test is similar to the test
performed by
\citeauthor{Plazar2019}.

To test our sampler, we sample $N$ solutions.
Let $S$ be the multiset that contains the sampled solutions such that $|S| = N$.
We then perform a frequency test for each variable.
We ignore variables that are always true or always false
because any sampler would generate the correct distribution for these variables
as long as the samples are models of the formula under test.

To test the individual variable frequencies
we perform Pearson's $\chi^2$ test on the vectors
$O$ and $E$ defined as follows.
We define
$O_{v=1} = |\{m \in S | m(v) = \textit{true}\}|$
(resp. $O_{v=0} = |\{m \in S | m(v) = \textit{false}\}|$) as the number of sampled
solutions with the target variable $v$ set to true (resp. false).
% We define $E_0$ and $E_1$ in a similar fashion.
We define
$E_{v=1} = \dfrac{N}{|R_F|} |\{m \in R_F | m(v) = \textit{true}\}|$
(resp.
$E_{v=0} = \dfrac{N}{|R_F|} |\{m \in R_F | m(v) = \textit{false}\}|$)
as the expected number of solutions
with the target variable set to true (resp. false)
under $H_0$.

We compute Pearson's $\chi^2$ test statistic for each variable as follows:
\begin{equation}
\chi_v^2 = \sum_{i \in \{1, 0\}} \dfrac{(O_{v=i} - E_{v=i})^2}{E_{v=i}}
\end{equation}

We follow by deriving individual p-values $p_v$ from each one-sided test
(with one degree of freedom).
Unfortunately, a direct comparison
with a significance level $\alpha$ is impossible, as performing multiple tests
raises the family-wise error rate (FWER), i.e.
performing multiple tests raises the probability of rejecting $H_0$
even though $H_0$ is true (type I error).
To mitigate this, we use the harmonic mean p-value defined in Subsection \ref{subs:meanp}.
We thus compute all the p-values $p_v$ associated with every variable (that
satisfies $E_{v=0} \neq 0 \wedge E_{v=1} \neq N$) and compute $\mathring{p}$
on the entire set of p-values (we give equal weights to all the p-values).
We reject $H_0$ if we find $\mathring{p} \leq \alpha$.

One may wish to ignore a variable $v$ if $E_{v=1}$
or $E_{v=0}$ is very small, as this will increase the
required sample size.
The reason is that
the approximation to the $\chi^2$ distribution in the GOF
test is known to break down if any expected number of occurrences
is below five \citep{Yates1934ContingencyTI}.
Thus, if we want $E_{v=i} \ge 5$, then we need to adjust $N$
accordingly. If one variable $v$ is true (resp. false) in most
models in $R_F$, then $N$ will increase. If the computational
cost is of importance, then we may wish to ignore some variables
with very low or very high frequencies.

Because we will have to repeat the VF test by using multiple formulae as input,
we ask if we may compute the HMP of HMPs.
By inserting the HMP formula into itself,
we find:

\begin{equation} \label{eq:hmp-hmp-p1}
\begin{split} 
    \mathring{p}_R
    &= \dfrac{\sum_{i \in R} w_i}{\sum_{i \in R} \dfrac{w_i}{\mathring{p}_{R_i}}} \\
    &= \dfrac
        {\sum_{i \in R} w_i}
        {\sum_{i \in R} \dfrac
            {w_i}
            {\dfrac
                {\sum_{j \in R_i} w_{ij}}
                {\sum_{j \in R_i} \dfrac{w_{ij}}{p_{ij}}}}} \\
\end{split}
\end{equation}

In our case, because we compute the HMP over all the individual monobit tests,
we know that $\forall R_i \; (\sum_{j \in R_i} w_{ij} = 1)$.
We thus obtain:

\begin{equation} \label{eq:hmp-hmp-p2}
\begin{split} 
    \mathring{p}_R
    &= \dfrac
        {\sum_{i \in R} w_i}
        {\sum_{i \in R} \dfrac
            {w_i}
            {\dfrac
                {1}
                {\sum_{j \in R_i} \dfrac{w_{ij}}{p_{ij}}}}} \\
    % &= \dfrac
    %     {\sum_{i \in R} w_i}
    %     {\sum_{i \in R} w_i \sum_{j \in R_i} \dfrac{w_{ij}}{p_{ij}}} \\
    % &= \dfrac
    %     {\sum_{i \in R} w_i}
    %     {\sum_{i \in R} \sum_{j \in R_i} \dfrac{w_i w_{ij}}{p_{ij}}} \\
    &= \dfrac
        {\sum_{i \in R} w_i \sum_{j \in R_i} w_{ij}}
        {\sum_{i \in R} \sum_{j \in R_i} \dfrac{w_i w_{ij}}{p_{ij}}} \\
    % &= \dfrac
    %     {\sum_{i \in R} \sum_{j \in R_i} w_i w_{ij}}
    %     {\sum_{i \in R} \sum_{j \in R_i} \dfrac{w_i w_{ij}}{p_{ij}}} \\
    &= \dfrac
        {\sum_{i \in R, j \in R_i} w_i w_{ij}}
        {\sum_{i \in R, j \in R_i} \dfrac{w_i w_{ij}}{p_{ij}}} \\
\end{split}
\end{equation}

We conclude that in our case, computing the HMP of HMPs is correct, as it is
equivalent to computing the HMP of the original p-values.

\subsubsection{Selected Features per Configuration (SFpC) Statistical Test}

\citet{DBLP:conf/splc/HeradioFGB20}
proposed a statistical test called the selected features
per configuration test (SFpC).

The main idea of the test is that a formula $F$ will have $n_k$
solutions with exactly $k$ variables set to true and the remaining
$|Var(F)| - k$ variables set to false with $n_k$
defined as follows
\begin{equation}
n_k = |\{m \in R_F | \; k = \sigma(m) \}|
\end{equation}
\begin{equation}
\sigma(m) = |\{v \in Var(F) | m(v) = \textit{true} \}|
\end{equation}
$n_k$ can be computed for every $0 \leq k \leq |Var(F)|$
with $k \in \mathbb{N}$. If we compute
$\dfrac{n_k}{|R_F|}$ for every $k$, we obtain the expected discrete
probability distribution, i.e. the distribution that
should be mimicked by the sampler if the sampler
is uniform.

Testing a sampler is performed as follows.
Let $S$ be the multiset containing the sampled solutions such that $|S| = N$.
The authors follow by defining
$E_i = n_i \dfrac{N}{|R_F|}$
, the expected number of solutions with exactly
$i$ variables set to true
and
$O_i = |\{m \in S | \; \sigma(m) = i \}|$
, the number of observed solutions in the sample $S$
with exactly $i$ variables set to true.

We compute Person's $\chi^2$ test statistic as follows:
\begin{equation}
\chi^2 = \sum_{i \in \Gamma} \dfrac{(O_{i} - E_{i})^2}{E_{i}}
\end{equation}

with $\Gamma = \{k \in \mathbb{N} | \; 0 \leq k \leq |Var(F)| \; \wedge E_k \neq 0\}$.
A p-value $p$ is derived by performing Pearson's one-sided $\chi^2$ test
(with $|\Gamma| - 1$ degrees of freedom).
We reject $H_0$ if we find $p \leq \alpha$
with $\alpha$, a predefined significance level.

The SFpC test uses a one-sided $\chi^2$ test to derive a p-value.
Our implementation differs from the original implementation in \citep{DBLP:conf/splc/HeradioFGB20}
(by using a one-sided Pearson's $\chi^2$ test) because we could not run their code.

\subsubsection{Birthday Problem Statistical Test}

% written by Martin. Please adapt the vocabulary that relates the application. But it would be better to not change the statistical vocabulary.

We finish by introducing a test inspired by
the birthday paradox and adapted from an existing
test on PRNGs \citep{ONeillbday}. Although the test is not necessarily
considered to be the most reliable within the PRNG community,
it does have some advantages. First, the test is simple enough
to adapt to URS as the required knowledge is limited to the
sample size and the total number of solutions for the
input Boolean formula. Second, it shows us whether
a sampler produced either too many or not enough duplicates
during the sampling process. If too many duplicates are produced, then
the sample may be of little value as the number of unique samples is low.
If too few duplicates have been produced, then the sample may not be uniform
but at least it likely explores a larger portion of the hyperspace.
We derive below the statistical test from the classical birthday problem \citep{blitzsteinIntroductionProbabilitySecond2019} to test the hypothesis that a sampler samples uniformly at random.

The statistics considered, noted $\mathrel{R}$, is the number of repeated solutions
, i.e.,
the number of sampled unordered pairs where both solutions are equal. 
The distribution of the statistic $\mathrel{R}$ under the null hypothesis 
(uniform random sampling) is a well-known result of the ``birthday problem''. 
The number of repeated pairs $R$ has approximately a Poisson distribution: 
$ \mathrel{R} \dot\sim \operatorname{Pois}(\lambda) $ 
with $\lambda = \dfrac{\binom{N}{2}}{|R_F|}$, $N$ the number of sampled 
solutions \cite[p.~179]{blitzsteinIntroductionProbabilitySecond2019}. 
Empirically, we sample $N$ samples from the sampler under test to compute 
the observed number of repeated samples noted $r$. We consider a two-sided 
test here, since our alternative hypothesis can imply a deviation from 
uniform sampling by the right (higher number of repeated samples) or 
by the left (lower number of repeated samples). The p-value of our 
two-sided birthday problem statistical test can be computed as follows:

\begin{equation} \label{eq:pvalue-birthday-test}
\begin{split} 
    p &= 2 \min \{ \mathbb{P}(\mathrel{R} \geq r \mid H_0 ), \, \mathbb{P}(\mathrel{R} \leq r \mid H_0 ) \} \\
      &= 2 \min \{ 1 - F_{\operatorname{Pois}(\lambda)}(r-1), \, F_{\operatorname{Pois}(\lambda)}(r) \} \;
\end{split}
\end{equation}

\noindent with $F_{\operatorname{Pois}(\lambda)}(r)$ the cumulative distribution function (CDF)
of the Poisson distribution parameterized by $\lambda$. Then, the p-value of our
birthday problem statistical test allows us to reject or not the null hypothesis.

\subsection{Combining Results from Multiple Formulae}

In contrast to testing pseudo-random number generators (PRNGs),
testing samplers also requires a formula to be given as input to
the sampler under test.
To alleviate possible biases due to the input formula,
we apply each statistical hypothesis test (aka statistical test) described above to multiple formulae. Unless otherwise stated, by the term \emph{test} we refer to the particular application of a given statistical test to a given sampler using a given Boolean formula. Thus, for a given statistical test and sampler, we conduct one test per available formula.

Each test produces a single p-value which can be used to make
inferences.
The family-wise error rate (FWER)
is the probability of making at least one type I error when performing
multiple tests.
The probability $\alpha$ of making a type I error
is the probability of rejecting the null hypothesis $H_0$
when $H_0$ is true.
Or in our case, the probability of wrongly rejecting the
uniformity of a sampler even though the sampler is uniform.

The probability of making a type I error if we perform
$N$ tests is computed as follows:
$$\alpha' = 1 - (1 - \alpha)^N$$
with $\alpha$ the significance level of a single test.
Thus, $\alpha'$ increases as the number of tests $N$ (in our case
formulae) increases.
This is undesirable as it may lead us to wrong conclusions
about the sampler under test.
If we set $\alpha = 0.01$ and $N = 20$ we find $\alpha'=0.18$.
This means that we have a probability of $0.18$ (if $H_0$
is true)
that at least one test will reject the null hypothesis
which may lead us to a wrong conclusion.
% Considering that according to our results, performing
% the tests multiple times with multiple formulae is necessary
% (and the fact that the VF test performs multiple tests itself),
% We need to address this issue considering our research questions.
Considering our research questions, this issue needs to be addressed.
There exists a multitude of controlling procedures to alleviate
this issue.

\subsubsection{Bonferroni Correction}

The Bonferroni correction
is a classical way of controlling
the FWER.
If $N$ tests are performed and we want to ensure
a probability of performing a type I error
of at most $\alpha$, then
we set $\alpha_i = \frac{\alpha}{N}$
with $\alpha_i$ the significance level of the individual tests.
The Bonferroni correction thus ensures that
$\alpha' = 1 - (1 - \alpha_i)^N \leq \alpha$
without imposing any assumptions about the dependence between the p-values.

\subsubsection{Harmonic Mean p-value (HMP)}
\label{subs:meanp}

The harmonic mean p-value (HMP) \citep{Good1958SignificanceTI,Wilson2019TheHM}
assesses the significance of groups
of hypotheses tests while also controlling the strong-sense
family-wise error rate. The Harmonic mean p-value improves on the
power of the Bonferroni correction.
Additionally, the HMP does not require the p-values to be independent.

The harmonic mean p-value is defined as follows:

\begin{equation}
\mathring{p}_R = \dfrac{
    \sum_{i \in R} w_i
}{
    \sum_{i \in R} \dfrac{w_i}{p_i}
}
\end{equation}

With $R$ any subset of the $m$ tests, $w_i$ the weights of each test,
$\sum_{i = 1}^m w_i = 1$ and, $p_i$
the p-value of each test.
To perform a family of tests at a significance level of approximately $\alpha$,
we reject the null hypothesis that none of the p-values
in the subset $R$ are significant when $\mathring{p}_R \leq \alpha w_R$
(with $w_R = \sum_{i \in R} w_i$).
The approximation is reasonable for a small significance level
$\alpha$ and improves with smaller values.
In other words, if we find $\mathring{p}_R \leq \alpha w_R$
then we can safely assume that at least one of the tests was significant
and thus reject the null hypothesis with a significance level of
approximately $\alpha$.

In this paper, we use the HMP.
We perform the desired statistical test on a set of formulae on a given sampler.
This gives us a vector of p-values on which we compute the HMP $\mathring{p}$.
% similarly to what is described in section \ref{subs:vfst}.
We may then compare $\mathring{p}$ to the appropriate significance level $\alpha$.
If we find $\mathring{p} \leq \alpha$ (because we compute $\mathring{p}$
on all the p-values we have $w_R = 1$),
then we know that at least one of the tests
is significant and we can reject $H_0$. Otherwise, we fail to reject $H_0$.
% and reject or fail to reject $H_0$.

\section{Experimental Study}
\label{sec:exp}
We define below our research questions and the general experimental settings common to all our experiments. The specific settings of each research question are detailed in Section~\ref{sec:results}.

\subsection{Research Questions}

Our study aims to answer the following research questions:
\begin{enumerate}
    \item \textbf{RQ1: Which URS tools are uniform according to the different statistical tests?}
    We assess the uniformity of the current implementations of state-of-the-art sampling methods. To achieve this, we apply separately our proposed statistical tests over a set of formulae and report the results for each statistical test and sampler.
    % \item \textbf{RQ2: Which tests are worth the computational cost?}
    \item \textbf{RQ2: What is the execution time of each statistical test?}
    Statistical tests have differences in reliability, meaning that the statistical values they focus on differ, leading to distinct abilities to detect non-uniformity. We therefore measure the execution time of each test on all samplers and input formulae. We contrast these measurements with the ability of each test to conclude non-uniformity.
    % Given the large configuration spaces, the statistical
    % tests only test for some aspects of uniformity (e.g. the variable frequencies
    % are an approximation of the true frequencies). Thus, a test's reliability
    % is limited by the test statistic and by the information used
    % to compute the test statistic.
    % Thus if we were to rely on a single test, designing
    % a sampler to induce wrong results would be easier.
    % However, testing some aspects of uniformity may require smaller sample
    % sizes than others. In other words, not every test has the same abilities
    % or the same computational cost.
    % We thus ask which tests are reliable enough to detect
    % non-uniformity while maintaining scalability.
    \item \textbf{RQ3: How do different formula datasets compare in non-uniformity detection ability?}
    We hypothesize that statistical test outcomes are strongly bound to the formulae used and that this can be a source of biases in uniformity test results. Specifically, we investigate to what extent synthetic formulae -- which are simpler and faster to process -- are useful to conclude about samplers' uniformity on real-world formulae.
    
    % A sampler being uniform on a formula
    % does not necessarily mean it is uniform on any formula.
    % Thus, we ask the question:
    % % Can we detect non-uniformity regardless of the
    % % input formula?
    % Will our statistical tests return the same answer regardless
    % of the formula given as input to the sampler under test?
    % A positive answer would lead to more trust in the test results
    % and allow us to use simpler formulae thus decreasing the computational cost.
    % However, a negative answer would indicate that the datasets
    % need to be chosen carefully.
\end{enumerate}

\subsection{Datasets}

We use a large number of well-known and publicly available models in our study, which are of various complexity and are either feature models or general Boolean formulae.

\subsubsection{Feature model benchmark properties} In total, we use the feature models of 128 real-world configurable systems (Linux, eCos, toybox, JHipster, etc.) with varying sizes and complexity. 
We first rely on 117 feature models used in ~\citep{DBLP:conf/sigsoft/KnuppelTMMS17, DBLP:conf/icse/KrieterTSSS18}. Most feature models contain between 1,221
and 1,266 features. Of these 117 models, 107 comprise between 2,968
and 4,138 cross-tree constraints, while one has 14,295 and the other
nine have between 49,770 and 50,606 cross-tree constraints~\citep{DBLP:conf/sigsoft/KnuppelTMMS17, DBLP:conf/icse/KrieterTSSS18}. 
Second, we include 10 additional feature models used in~\citep{Liang:2015:SAL:2791060.2791070} and not in~\citep{DBLP:conf/sigsoft/KnuppelTMMS17, DBLP:conf/icse/KrieterTSSS18}; they also contain a large number of features (e.g., more than 6,000).  
Third, we also add the JHipster feature model~\citep{raible:jhipsterBook, Halin2018} to the study, a realistic but relatively smaller feature model (45 variables, 26,000+ configurations). We later refer to these benchmarks as the feature model benchmark. Once put in conjunctive normal form, these instances typically contain between 1 and 15 thousand variables and up to 340 thousand clauses. The hardest of them, modeling the Linux kernel configuration, contains more than 6 thousand variables, and 340 thousand clauses, and is generally seen as a milestone in configurable system analysis.

\subsubsection{General Boolean formulae}

In addition to these feature models,
we use the industrial SAT formulae as used in
\citep{Dutra2018}.
Since these formulae are much smaller than the feature
models we use (typically a few thousand clauses), they will
provide a basis of results for statistical analysis, in case a solver
cannot produce enough samples on the harder formulae.
We later refer to these benchmarks as the non-feature model
benchmarks.

\subsubsection{Filtered dataset $\Omega$}

Performing tests on samplers requires us to generate a large number of samples.
Thus, to limit the computation time we combine both the feature model dataset and the general Boolean
formulae. We then ran the UniGen3 sampler on all of these formulae and measured the time and
memory UniGen3 required to sample 1000 samples. We removed all the formulae that required more
than 10 minutes or more than 400MB of memory, which left us with 195 formulae.
In the following sections, we will refer to this dataset as the $\Omega$ dataset.

We chose UniGen3 as a reference sampler as it has theoretical guarantees
and it is the slowest sampler on our list according to our
experiments (if we exclude Smarch).
Given how slow Smarch was in our experiments, relying on Smarch to filter
the dataset would have left us with a smaller dataset.

\subsubsection{Synthetic formulae}

To test the importance of a dataset, we also generated synthetic k-CNF formulae.
To generate clauses in k-CNF formulae, we select k unique
variables at random and negate them
with probability $\frac{1}{2}$.

The first synthetic dataset is r30c90 which contains 300 satisfiable 3-CNF formulae
consisting of 30 variables and 90 clauses.
The second synthetic dataset is r30c114, which contains 300 satisfiable 3-CNF formulae
consisting of 30 variables and 114 clauses.

We also generated a dataset named r30c150b1000 which contains 300 3-CNF formulae.
Each formula is generated with 30 variables and 150 clauses.
We then followed by generating 1000
assignments to the 30 variables and negating the literals in the clauses such that
the generated assignment is a solution to the formula \citep{Escamocher2022GenerationAP}.

\subsection{Infrastructure}

The experiments
% regarding the computation of the equivalence classes,
% the MIS computation as well as the time and memory usages of the samplers
were computed on an HPC containing 354 nodes, each of which has
256 GB of RAM and 2 AMD Epyc ROME 7H12 CPUs running at 2.6 GHz.

\subsection{Computation budget}

We performed each statistical test on each sampler and each dataset.
We define a unit as one statistical test, one sampler, and one dataset.
Each unit had two days on a full node of the HPC. If the test was not finished
by then, it would be interrupted, and we used the available results.

As we require large amounts of samples (and because some samplers
are designed to generate no duplicate samplers even though some statistical
tests require them), we decided to generate samples in batches.
This means that we repeatedly called the samplers asking for
1000 samples until the required number of samples was met.
The sampling process had a timeout of 5 hours.
This means that if sampling the required amount of samples would take longer
than 5 hours, then the test is not performed on the input formula
(but may be performed on other formulae within the unit).

\section{Results}
\label{sec:results}
\subsection{RQ1: Uniformity of samplers}

In this section, we present the general results regarding uniformity
performed on our $\Omega$ dataset.
Table \ref{tab:real} reports the results.
For each test, we split them into two columns,
the first one is the number of formulae on which we managed to
perform the test, and the second is the
% uniformity result. A 'T' means that the sampler sampled uniformly
% according to the test and an empty cell means that the sampler
% did not sample uniformly according to the test.
p-value. A bold p-value indicates that the p-value
is greater than our predefined significance level $\alpha = 0.01$
(and thus is uniform according to our test).
% A '-' indicates that we did not manage to perform the test
% on any formula.
The tests were not necessarily performed on all the formulae
because the sampler may have crashed or the sampling process
may have been too time-consuming and thus interrupted before
it could terminate incidentally canceling the test.

We observe that almost all samplers fail the monobit test
on the $\Omega$ dataset, except KUS, SPUR, and UniGen3.
The only sampler that succeeds in the variable frequency test is UniGen3.
This result is of particular interest as SPUR has been deemed non-uniform by our test
and SPUR is used as a reference sampler in Barbarik \citep{MPC20}.
Moreover, our results indicate that UniGen3
is likely to be the only uniform sampler.
We note that the SFpC test does not seem to give any additional information
as the VF test has already detected all the non-uniform samplers.
The birthday test fails to detect the non-uniformity of Smarch,
while all our other tests do.
Our results indicate that the variable frequency test is of sufficient
reliability to detect non-uniform samplers. The SFpC test can be used
to confirm the results of the VF test.
The Monobit and birthday tests are decent tests but
seem less reliable, i.e., the tests may return that the sampler is uniform
even though it is not. In other words, their use requires the use
of other tests such as the VF test to confirm the uniformity results.

We note that the samplers KUS and STS
passed the GOF test.
This is somewhat surprising, as it contradicts our other results.
However, given that the GOF test was performed successfully on a much lower
number of formulae, we argue that the test is less reliable
and conclude that only UniGen3 is uniform.

\begin{table}
\makebox[\textwidth]{%
    \begin{tabular}{c|c|c|c|c|c|c|c|c|c|c}
& \multicolumn{2}{c|}{Monobit}& \multicolumn{2}{c|}{VF}& \multicolumn{2}{c|}{Birthday}& \multicolumn{2}{c|}{SFpC}& \multicolumn{2}{c}{GOF} \\
% sampler & \#F & Uniform? & \#F & Uniform? & \#F & Uniform? & \#F & Uniform? & \#F & Uniform?\\
sampler & \#F & p-value & \#F & p-value & \#F & p-value & \#F & p-value & \#F & p-value\\
\hline
% KUS          &  193 & T &  192 &   &  146 &   &   86 &   &   27 & T \\
% QuickSampler &  188 &   &  186 &   &  139 &   &   76 &   &   31 &   \\
% Smarch       &  178 &   &  104 &   &   65 & T &   28 &   &    3 &   \\
% SPUR         &  193 & T &  189 &   &  140 &   &   99 &   &   44 &   \\
% STS          &  193 &   &  191 &   &  150 &   &   80 &   &   30 & T \\
% CMSGen       &  144 &   &  143 &   &  100 &   &   71 &   &   46 &   \\
% UniGen3      &  192 & T &  183 & T &  117 & T &   74 & T &   44 & T \\

KUS          &  193 & \textbf{0.164} &  192 & 0.000 &  146 & 0.001 &   86 & 0.000 &   27 & \textbf{0.197} \\
QuickSampler &  188 & 0.000 &  186 & 0.000 &  140 & 0.000 &   76 & 0.000 &   31 & 0.000 \\
Smarch       &  178 & 0.000 &  104 & 0.000 &   58 & \textbf{0.107} &   28 & 0.000 &    3 & 0.002 \\
SPUR         &  193 & \textbf{0.073} &  189 & 0.000 &  140 & 0.000 &   99 & 0.000 &   44 & 0.000 \\
STS          &  193 & 0.000 &  191 & 0.000 &  150 & 0.000 &   80 & 0.000 &   30 & \textbf{1.000} \\
CMSGen       &  144 & 0.000 &  143 & 0.000 &  100 & 0.000 &   71 & 0.000 &   46 & 0.000 \\
UniGen3      &  192 & \textbf{0.229} &  183 & \textbf{0.104} &  116 & \textbf{0.278} &   74 & \textbf{0.180} &   44 & \textbf{0.293} \\

    \end{tabular}}
    \caption{Experimental results for the $\Omega$ dataset.
        For each test (and for each formula), each sampler was called
        multiple times to generate samples of size 1000. The bold p-values
        are all greater than our significance level $\alpha = 0.01$.
        \#F indicates the number of formulae on which the test was successfully
        performed (i.e. without timeouts or out-of-memory errors).}
    \label{tab:real}
\end{table}

% \subsubsection{Birthday test results}

As noted above, most samplers are not uniform. Moreover, it is well-known that
there exist formulae for which uniform sampling does not scale.
Thus, heuristic-based samplers become interesting. However, not all heuristic-based
samplers are equal.

The birthday test provides very insightful information
on samplers (in addition to the p-value),
as the birthday test studies the number of repetitions.
The observed number of repetitions may be of particular interest if the user wishes to
use a heuristic-based sampler.
A sampler with too many repetitions is non-uniform and often returns
the same solution. A sampler with too few repetitions is non-uniform and
seldom returns the same solution. The first case would lead to
the exploration of a smaller part of the solution space.
On the other hand, generating fewer repetitions than necessary for uniformity
on large solution spaces would indicate a better exploration of said
solution space.
Suppose that a user wishes to sample uniformly from a formula
but none of the theoretically uniform tools fit into the user's budget.
The user thus has the choice between QuickSampler, CMSGen, and STS.
The user may sample from the target formula by using all the samplers
and comparing the observed numbers of repetitions. The sampler generating
the fewest repetitions is likely a good candidate.

\begin{table}
    \begin{tabular}{c|c|c|c|c|c|c}
    & \multicolumn{2}{c|}{Uniformity}& \multicolumn{4}{c}{Observed number of repetitions} \\
% sampler & \#F & Uniform? & min & max & average & median\\
sampler & \#F & p-value & min & max & average & median\\
\hline
% KUS          &  146 &   &       0 &      17 &    8.42 &       9 \\
% QuickSampler &  139 &   &       0 &   20974 &  356.04 &       4 \\
% Smarch       &   65 & T &       4 &      16 &    9.28 &       9 \\
% SPUR         &  140 &   &       3 &      63 &   17.74 &      12 \\
% STS          &  150 &   &       0 &    1036 &   16.24 &       6 \\
% CMSGen       &  100 &   &       5 &  129188 & 2927.82 &      32 \\
% UniGen3      &  117 & T &       4 &      22 &    9.79 &      10 \\

KUS          &  146 & 0.001 &       0 &      18 &    8.76 &       9 \\
QuickSampler &  140 & 0.000 &       0 &   29873 &  487.99 &       5 \\
Smarch       &   58 & \textbf{0.107} &       4 &      21 &   10.88 &      10 \\
SPUR         &  140 & 0.000 &       3 &     306 &   34.09 &      12 \\
STS          &  150 & 0.000 &       0 &    1055 &   16.09 &       6 \\
CMSGen       &  100 & 0.000 &       6 & 3846964 & 77232.35 &      32 \\
UniGen3      &  116 & \textbf{0.278} &       4 &      19 &   10.23 &      10 \\
    \end{tabular}
    \caption{Extended experimental results for the
        birthday test with the $\Omega$ dataset.
        For each formula, each sampler was called
        multiple times to generate samples of size 1000. The bold p-values
        are all greater than our significance level $\alpha = 0.01$.
        \#F indicates the number of formulae on which the test was successfully
        performed (i.e. without timeouts or out-of-memory errors).}
    \label{tab:bday:real}
\end{table}

Table \ref{tab:bday:real} shows the detailed results for the Birthday
test on our $\Omega$ dataset.
In our experiments, we set the expected number of repetitions
to be approximately 10 as shown by the average number
of repetitions observed on the samples generated by UniGen3 (10.23).
The CMSGen sampler generated on average 77232.35 repetitions. This number is likely
too high for most users. The same is true for QuickSampler which generated an average of
487.99 repetitions.
STS on the other hand generated an average of 16.09 repetitions, which is too high to be
uniform but much lower than QuickSampler and CMSGen.
Thus, a user who cannot afford uniform sampling should likely rely on STS
for their sampling tasks.

\begin{framed}
    \textbf{Answer to RQ1:}
    According to our results, the monobit test is not very reliable.
    The variable frequency test detected the non-uniformity of every
    sampler, except for UniGen3.
    The SFpC test did not detect
    deviations from uniformity that the VF test did not already detect.
    Thus, SFpC test does not seem to
    add more information than the variable
    frequency test but may become useful in the future as more samplers
    are developed.
    The GOF test requires too many samples to be usable,
    given the computational cost of sampling.
    
    The birthday test is an interesting addition to the set of tests.
    It allows for uniformity testing and provides further insight into
    the capacity of samplers to explore the solution space.
\end{framed}

\subsection{RQ2: Scalability}

\begin{table}
\makebox[\textwidth]{%
    \begin{tabular}{c|c|c|c|c|c|c|c|c|c|c}
& \multicolumn{2}{c|}{Monobit}& \multicolumn{2}{c|}{VF}& \multicolumn{2}{c|}{Birthday}& \multicolumn{2}{c|}{SFpC}& \multicolumn{2}{c}{GOF} \\
sampler & \#F & time (h) & \#F & time (h) & \#F & time (h) & \#F & time (h) & \#F & time (h)\\
\hline
KUS          &  193 &   0.8 &  192 &   7.6 &  146 &   8.0 &   86 &  43.6 &   27 &   0.2 \\
QuickSampler &  188 &   4.1 &  186 &  17.2 &  140 &  21.5 &   76 &  69.7 &   31 &   2.3 \\
Smarch       &  178 & 401.9 &  104 & 196.8 &   58 &  84.0 &   28 &  36.2 &    3 &   0.0 \\
SPUR         &  193 &   4.1 &  189 &   4.2 &  140 &   1.5 &   99 &  45.0 &   44 &   1.1 \\
STS          &  193 &   8.3 &  191 &  27.4 &  150 &  34.4 &   80 &  38.3 &   30 &   0.3 \\
CMSGen       &  144 &   0.1 &  143 &   0.9 &  100 &   0.5 &   71 &  17.2 &   46 &   1.1 \\
UniGen3      &  192 &   2.6 &  183 &  31.4 &  116 &   9.2 &   74 &  36.5 &   44 &   2.4 \\

    \end{tabular}}
    \caption{Scalability results for the $\Omega$ dataset.
        For each test (and for each formula), each sampler was called
        multiple times to generate samples of size 1000.
        The indicated time (in hours) is the accumulated time across all the formulae
        for which the test was performed successfully.
        \#F indicates the number of formulae on which the test was successfully
        performed (i.e. without timeouts or out-of-memory errors).}
    \label{tab:t:real}
\end{table}

In this subsection, we explore the scalability of each test, or in other words,
the computational cost of each statistical test presented in this paper.

Table \ref{tab:t:real} shows the sequential execution time of
each test for the $\Omega$ dataset on each sampler.
The sequential execution time is computed on the indicated number of formulae
(and it does not contain the formulae that timed out or had an error).
A first striking observation
is that Smarch is significantly slower than all the other samplers.
A second more subtle observation is that the execution time
increases when switching from the monobit test to the VF test
and from the VF test to the SFpC test.
At the same time, the number of formulae on which the tests
were performed successfully gradually decreases from one test to another.
This indicates that the SFpC test requires more computation time than the VF test,
which requires more computation time than the monobit test.
The birthday test is more nuanced regarding the accumulated computation time.
However, regarding the number of formulae on which the birthday test was
performed successfully, we find that the birthday test is more expensive than the VF
test and cheaper than the SFpC test.
The GOF test was successfully performed on the smallest number of formulae.
We thus conclude that GOF is the most expensive test to perform.

Given our results, if we order the test according to their computational cost,
we find the following ordering:
Monobit $\leq_C$
VF $\leq_C$
Birthday $\leq_C$
SFpC $\leq_C$
GOF,
with monobit the cheapest test and GOF the most expensive test.

Smarch seems to observe a decrease in execution time
when reading the table according to our ordering.
However, this is coupled with a drastic decrease
in the number of formulae successfully processed.

% \zyno{add discussion regarding the information given by each test
% to further justify the test ordering??}

\begin{figure}
    \centering
    \begin{tikzpicture}
        \node (s1) {\textbf{start}};
        
        \node (vf) [right=of s1] {VF};
        \node (m) [above=of vf] {Monobit};
        \node (b) [below=of vf] {Birthday};
        \node (sfpc) [right=of vf] {SFpC};
        \node (gof) [right=of sfpc] {GOF};

        \draw[->,dashed] (s1) -- (m);
        \draw[->] (s1) -- (vf);
        \draw[->] (m) -- (vf);
        \draw[->] (vf) -- (sfpc);
        \draw[->,dashed] (vf) -- (b);
        \draw[->] (b) -- (sfpc);
        \draw[->,dashed] (sfpc) -- (gof);
        
    \end{tikzpicture}
    \caption{Statistical test ordering for uniform random sampling. The dashed lines indicate optional transitions. A user may thus adapt the executed tests depending on their needs and computational budget.}
    \label{fig:ord}
\end{figure}
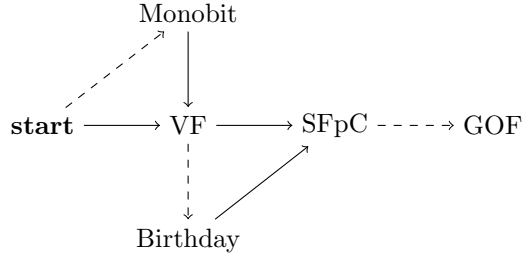

By considering our results on scalability and uniformity,
we propose the ordering presented in Figure \ref{fig:ord}.
The dashed lines indicate optional transitions.
A user with computational budget limitations
would start testing with the VF test and then proceed with
the SFpC test. If the user has a bigger computational budget,
then they may include the monobit test before the VF test
or the birthday test after the VF test. The choice depends
on the user's requirements. If the user wants to analyze the observed
number of repetitions then the birthday test is interesting.
However, if the user has many non-uniform samplers, then the monobit
test may be a good candidate for fast 'pre-testing'.
Additionally, the choice depends on the user's computational budget as
the monobit test is cheaper than the birthday test,
The GOF test is optional as it only scales to the smallest formulae.

\begin{framed}
    \textbf{Answer to RQ2:}
    After an analysis of the computational cost of each test,
    we find the following ordering:
    Monobit $\leq_C$
    VF $\leq_C$
    Birthday $\leq_C$
    SFpC $\leq_C$
    GOF,
    with monobit the cheapest test and GOF the most expensive test.

    We recommend performing the tests
    according to the ordering in Figure \ref{fig:ord}.
    If performing every test is too expensive we recommend
    excluding the GOF test first.
\end{framed}

\subsection{RQ3: On the influence of formula choice}

In this section, we explore the influence of the dataset on the uniformity
results.
The goal is to understand if the dataset influences
the uniformity result returned by our statistical test.
A negative answer would allow us to focus our testing on small formulae
that are easy to sample from. A positive answer would mean that
to obtain reliable results, one needs to choose their dataset carefully.

% In this section, we examine how the dataset affects the uniformity results.
% Our objective is to determine whether the dataset impacts the uniformity result produced by our statistical test.
% A negative answer would allow us to focus on small formulae that are easy to sample from.
% A positive answer suggests that careful dataset selection is necessary for reliable results.

To explore the influence of the dataset we also performed the statistical tests
by using synthetic formulae. We may then compare these results with
the results on real-world formulae.

\subsubsection{Uniformity}

\begin{table}
\makebox[\textwidth]{%
    \begin{tabular}{c|c|c|c|c|c|c|c|c|c|c}
& \multicolumn{2}{c|}{Monobit}& \multicolumn{2}{c|}{VF}& \multicolumn{2}{c|}{Birthday}& \multicolumn{2}{c|}{SFpC}& \multicolumn{2}{c}{GOF} \\
% sampler & \#F & Uniform? & \#F & Uniform? & \#F & Uniform? & \#F & Uniform? & \#F & Uniform?\\
sampler & \#F & p-value & \#F & p-value & \#F & p-value & \#F & p-value & \#F & p-value\\
\hline
% KUS          &  300 & T &  300 & T &  300 & T &  300 & T &  300 & T \\
% QuickSampler &  300 & T &  300 &   &  300 &   &  300 &   &  300 &   \\
% Smarch       &  300 & T &  292 & T &  300 & T &   27 & T &   21 & T \\
% SPUR         &  300 & T &  300 &   &  300 &   &  300 & T &  300 &   \\
% STS          &  300 & T &  300 &   &  300 &   &  300 & T &  300 & T \\
% CMSGen       &  300 &   &  300 &   &  300 &   &  300 &   &  300 &   \\
% UniGen3      &  300 & T &  300 & T &  300 & T &  300 & T &  300 & T \\

\hline
KUS          &  300 & \textbf{0.192} &  300 & \textbf{0.052} &  300 & \textbf{0.126} &  300 & \textbf{0.044} &  300 & \textbf{0.214} \\
QuickSampler &  300 & \textbf{0.824} &  300 & 0.000 &  300 & 0.000 &  300 & 0.000 &  300 & 0.000 \\
Smarch       &  300 & \textbf{0.071} &  292 & \textbf{0.083} &  300 & \textbf{0.186} &   27 & \textbf{0.103} &   21 & \textbf{0.060} \\
SPUR         &  300 & \textbf{0.046} &  300 & 0.002 &  300 & 0.000 &  300 & \textbf{0.206} &  300 & 0.000 \\
STS          &  300 & \textbf{0.666} &  300 & 0.000 &  300 & 0.000 &  300 & \textbf{0.012} &  300 & \textbf{1.000} \\
CMSGen       &  300 & 0.000 &  300 & 0.000 &  300 & 0.000 &  300 & 0.000 &  300 & 0.000 \\
UniGen3      &  300 & \textbf{0.061} &  300 & \textbf{0.077} &  300 & \textbf{0.128} &  300 & \textbf{0.072} &  300 & \textbf{0.123} \\
    \end{tabular}}
    \caption{Experimental results for the r30c90 dataset.
        For each test (and for each formula), each sampler was called
        multiple times to generate samples of size 1000. The bold p-values
        are all greater than our significance level $\alpha = 0.01$.
        \#F indicates the number of formulae on which the test was successfully
        performed (i.e. without timeouts or out-of-memory errors).}
        
    \label{tab:r30c90}
\end{table}

\begin{table}
\makebox[\textwidth]{%
    \begin{tabular}{c|c|c|c|c|c|c|c|c|c|c}
& \multicolumn{2}{c|}{Monobit}& \multicolumn{2}{c|}{VF}& \multicolumn{2}{c|}{Birthday}& \multicolumn{2}{c|}{SFpC}& \multicolumn{2}{c}{GOF} \\
% sampler & \#F & Uniform? & \#F & Uniform? & \#F & Uniform? & \#F & Uniform? & \#F & UnifoIrm?\\
sampler & \#F & p-value & \#F & p-value & \#F & p-value & \#F & p-value & \#F & p-value\\
\hline
% KUS          &  296 & T &  300 & T &  300 & T &  300 & T &  300 & T \\
% QuickSampler &  296 & T &  296 &   &  300 &   &  296 &   &  300 &   \\
% Smarch       &  296 &   &  296 &   &  296 & T &  284 &   &  275 &   \\
% SPUR         &  296 & T &  300 & T &  300 &   &  300 & T &  300 &   \\
% STS          &  296 & T &  300 &   &  300 &   &  300 & T &  300 & T \\
% CMSGen       &  296 &   &  300 &   &  300 &   &  300 &   &  300 &   \\
% UniGen3      &  296 & T &  300 & T &  300 & T &  300 & T &  300 & T \\

\hline
KUS          &  296 & \textbf{0.160} &  300 & \textbf{0.094} &  300 & \textbf{0.010} &  300 & \textbf{0.031} &  300 & \textbf{0.130} \\
QuickSampler &  296 & \textbf{0.958} &  296 & 0.000 &  300 & 0.000 &  296 & 0.000 &  300 & 0.000 \\
Smarch       &  296 & 0.000 &  296 & 0.000 &  296 & \textbf{0.150} &  284 & 0.000 &  275 & 0.001 \\
SPUR         &  296 & \textbf{0.189} &  300 & \textbf{0.125} &  300 & 0.000 &  300 & \textbf{0.092} &  300 & 0.000 \\
STS          &  296 & \textbf{0.974} &  300 & 0.000 &  300 & 0.000 &  300 & \textbf{0.924} &  300 & \textbf{1.000} \\
CMSGen       &  296 & 0.000 &  300 & 0.000 &  300 & 0.000 &  300 & 0.000 &  300 & 0.000 \\
UniGen3      &  296 & \textbf{0.198} &  300 & \textbf{0.155} &  300 & \textbf{0.198} &  300 & \textbf{0.028} &  300 & \textbf{0.066} \\
    \end{tabular}}
    \caption{Experimental results for the r30c114 dataset.
        For each test (and for each formula), each sampler was called
        multiple times to generate samples of size 1000. The bold p-values
        are all greater than our significance level $\alpha = 0.01$.
        \#F indicates the number of formulae on which the test was successfully
        performed (i.e. without timeouts or out-of-memory errors).}
    \label{tab:r30c114}
\end{table}

\begin{table}
\makebox[\textwidth]{%
    \begin{tabular}{c|c|c|c|c|c|c|c|c|c|c}
& \multicolumn{2}{c|}{Monobit}& \multicolumn{2}{c|}{VF}& \multicolumn{2}{c|}{Birthday}& \multicolumn{2}{c|}{SFpC}& \multicolumn{2}{c}{GOF} \\
% sampler & \#F & Uniform? & \#F & Uniform? & \#F & Uniform? & \#F & Uniform? & \#F & Uniform?\\
sampler & \#F & p-value & \#F & p-value & \#F & p-value & \#F & p-value & \#F & p-value\\
\hline
% KUS          &  300 & T &  300 & T &  300 & T &  300 & T &  300 & T \\
% QuickSampler &  300 & T &  300 &   &  300 &   &  300 &   &  300 &   \\
% Smarch       &  300 & T &  271 & T &  300 & T &   25 & T &   11 & T \\
% SPUR         &  300 & T &  300 & T &  300 &   &  300 & T &  300 &   \\
% STS          &  300 & T &  300 &   &  300 &   &  300 &   &  300 & T \\
% CMSGen       &  300 &   &  300 &   &  300 &   &  300 &   &  300 &   \\
% UniGen3      &  300 & T &  300 & T &  300 & T &  300 & T &  300 & T \\

\hline
KUS          &  300 & \textbf{0.197} &  300 & \textbf{0.099} &  300 & \textbf{0.242} &  300 & \textbf{0.101} &  300 & \textbf{0.073} \\
QuickSampler &  300 & \textbf{0.837} &  300 & 0.000 &  300 & 0.000 &  300 & 0.000 &  300 & 0.000 \\
Smarch       &  300 & \textbf{0.126} &  271 & \textbf{0.110} &  300 & \textbf{0.196} &   25 & \textbf{0.177} &   11 & \textbf{0.282} \\
SPUR         &  300 & \textbf{0.135} &  300 & \textbf{0.109} &  300 & 0.000 &  300 & \textbf{0.210} &  300 & 0.000 \\
STS          &  300 & \textbf{0.607} &  300 & 0.000 &  300 & 0.000 &  300 & 0.003 &  300 & \textbf{1.000} \\
CMSGen       &  300 & 0.001 &  300 & 0.000 &  300 & 0.000 &  300 & 0.000 &  300 & 0.000 \\
UniGen3      &  300 & \textbf{0.138} &  300 & \textbf{0.084} &  300 & \textbf{0.141} &  300 & \textbf{0.033} &  300 & \textbf{0.104} \\

    \end{tabular}}
    \caption{Experimental results for the r30c150b1000 dataset.
        For each test (and for each formula), each sampler was called
        multiple times to generate samples of size 1000. The bold p-values
        are all greater than our significance level $\alpha = 0.01$.
        \#F indicates the number of formulae on which the test was successfully
        performed (i.e. without timeouts or out-of-memory errors).}
    \label{tab:r30c150b1000}
\end{table}

We start by exploring the uniformity results obtained
by performing the statistical tests on synthetic datasets.
Tables \ref{tab:r30c90}, \ref{tab:r30c114}, and \ref{tab:r30c150b1000}
show the uniformity results on the synthetic datasets.
As with Table \ref{tab:real}
% 'T' indicates that the sampler
% is considered uniform, the absence of symbols indicates
% non-uniformity
a bold p-value indicates that the p-value
is greater than our predefined significance level $\alpha = 0.01$
(an is thus uniform according to our test).
% A '-' indicates that we did not manage to perform the test
% on any formula.

Our first observation is that KUS seems to be uniform on the synthetic
formulae as it fails no test.
We also observe that CMSGen fails the monobit test on every synthetic dataset.
This gives us a high confidence in the non-uniformity of CMSGen
as the monobit test seems to have little reliability when used with synthetic datasets.
Moreover, CMSGen fails every other test on every dataset and we thus conclude
that CMSGen is not uniform.

We follow by observing that the dataset used in Table \ref{tab:r30c90} is more effective
at detecting non-uniformity than the dataset used in Table \ref{tab:r30c114}.
However, the dataset r30c90 fails at detecting the non-uniformity of Smarch
with the VF test while it succeeds when performed by using the r30c114 dataset.
Thus to detect the non-uniformity of as many samplers as possible
by using the VF test and synthetic datasets we would need to combine
the results obtained with both datasets, r30c90 and r30c114.

We continue this subsection by
comparing the uniformity results with the ones found in Table \ref{tab:real}
which were computed on our $\Omega$ dataset.
This should give us insights into the importance of dataset choice.

We start by noting that the monobit test is significantly less reliable on synthetic formulae
as it eliminates fewer samplers than in Table \ref{tab:real}.
Next, our results on synthetic formulae indicate that KUS is uniform and
gives us mixed results on STS.
The results in Table \ref{tab:real} are much clearer as both STS and KUS fail
every test except for the GOF test (and the monobit test for KUS).
Thus, we conclude that the choice of the dataset is important for reliable
uniformity results.

\subsubsection{Scalability}

We continue this subsection by exploring the difference in scalability
between the real-world dataset and the synthetic datasets.
This will give us further insights regarding the relevance of synthetic datasets.

\begin{table}
\makebox[\textwidth]{%
    \begin{tabular}{c|c|c|c|c|c|c|c|c|c|c}
& \multicolumn{2}{c|}{Monobit}& \multicolumn{2}{c|}{VF}& \multicolumn{2}{c|}{Birthday}& \multicolumn{2}{c|}{SFpC}& \multicolumn{2}{c}{GOF} \\
sampler & \#F & time (h) & \#F & time (h) & \#F & time (h) & \#F & time (h) & \#F & time (h)\\
\hline
KUS          &  300 &   0.0 &  300 &   0.0 &  300 &   0.0 &  300 &   1.0 &  300 &   1.7 \\
QuickSampler &  300 &   0.2 &  300 &   0.3 &  300 &   0.1 &  300 &   3.9 &  300 &   4.9 \\
Smarch       &  300 & 321.7 &  292 & 369.6 &  300 & 453.1 &   27 &  68.2 &   21 &  54.1 \\
SPUR         &  300 &   0.0 &  300 &   0.0 &  300 &   0.0 &  300 &   0.3 &  300 &   0.5 \\
STS          &  300 &   0.0 &  300 &   0.0 &  300 &   0.0 &  300 &   0.3 &  300 &   0.5 \\
CMSGen       &  300 &   0.0 &  300 &   0.0 &  300 &   0.0 &  300 &   0.1 &  300 &   0.2 \\
UniGen3      &  300 &   0.0 &  300 &   0.1 &  300 &   0.1 &  300 &   1.7 &  300 &   2.5 \\
    \end{tabular}}
    \caption{Scalability results for the r30c90 dataset.
        For each test (and for each formula), each sampler was called
        multiple times to generate samples of size 1000.
        The indicated time (in hours) is the accumulated time across all the formulae
        for which the test was performed successfully.
        \#F indicates the number of formulae on which the test was successfully
        performed (i.e. without timeouts or out-of-memory errors).}
    \label{tab:t:r30c90}
\end{table}

Table \ref{tab:t:r30c90} shows the sequential execution time of
each test for the r30c90 dataset on each sampler.
We observe that except for the Smarch sampler,
the execution times are generally very low (under 1 hour).

By comparing Tables \ref{tab:t:real} and \ref{tab:t:r30c90}
we notice that the synthetic dataset is significantly faster to process
than the $\Omega$ dataset.
As an example, the VF test on STS required 27.4 hours to be performed on the $\Omega$
dataset and required around 0.02 hours to be performed on the r30c90 dataset.
The SFpC time budget for STS on the synthetic dataset was 0.3 hours which is
also less than required by the VF test on the $\Omega$ dataset
and also less than the monobit test on the $\Omega$ dataset (8.3 hours).
However, all of these tests have the same conclusion: STS is not uniform.
We thus argue that fast 'pre-testing' may be performed by using synthetic datasets
to detect the majority of non-uniform samplers quickly.
This would allow one to filter a lot of non-uniform samplers while testing
and only perform more expensive tests on more expensive datasets if a sampler
'survives' the 'pre-testing'.

\begin{framed}
    \textbf{Answer to RQ3:}
    We conclude that the dataset choice can drastically bias the uniformity conclusion. Specifically, if a sampler fails a test on a synthetic benchmark, then it is unlikely
    to pass the test on a real-world formula. The contraposition, however, does not hold. 
    Thus, we recommend using synthetic formulae to quickly eliminate
    a large number of samplers before considering larger, real-world formulae.
\end{framed}

\subsection{Discussion on uniformity and statistical test results}

Every statistical test presented in this article has different strengths
and weaknesses.
For example, the monobit test operates on two categories.
It is thus likely to require fewer samples as this means
that it is easier to have the required number of observations
in each category (unless there is a strong imbalance between
both categories). The low sample size is the monobit test's
biggest strength.
However, if the non-uniformity of a sampler does not generate an effect
across these two categories then the monobit test will not detect it,
thus the test's biggest weakness.
A sampler may be engineered to specifically pass the monobit
test.
Our results show that SPUR is uniform according to the monobit test
but not uniform according to the VF test, which further confirms our claim that
using a single statistical test gives unreliable results.
This is also the conclusion of the PRNG community as
it is standard practice in statistical testing
of pseudo-random number generators (PRNG) \citep{rukhin2001statistical}.

We observe from our results that formula choice has an impact on
test results. For example, KUS was detected as not uniform on the real-world
formulae while being uniform on the synthetic datasets.
This gives us two insights. First, testing a sampler on a single
formula is not enough. Second, the used dataset should contain a variety
of formulae to maximize the coverage of a sampler (similar to software
testing).

Our next observation is that in most cases a sampler (whether uniform or not)
will usually fail a test on at least a few formulae.
Statistical tests have a non-zero probability of returning the wrong answer
(i.e. Type-I error).
That is, sometimes a non-uniform sampler will pass a test, and sometimes
a uniform sampler will fail a test.
If a sampler is uniform, then the p-values returned by
the statistical tests will also have a uniform distribution \citep{Klammer2009}.
This means that a sampler failing a test on a single formula (if we test
with multiple formulae) is not enough to determine if the sampler is not
uniform.
This raises the question, how many failed tests are too many?
Luckily, the issue of performing multiple statistical tests and its
consequences has been studied. We use the HMP to mitigate these issues.
By using the HMP we can summarize the test results for a dataset
with a single p-value, thus simplifying the interpretation.

To conclude, we need to test our samplers by using multiple tests
and by using a diverse set of formulae to explore the uniformity of
the samplers as much as possible.
Performing multiple tests means that we may find a sampler which
will pass some tests and fail others.
To decide whether a sampler is uniform or not, we look at the consistency
of the results instead of the specific results.
If we test a sampler with five tests and the sampler fails three of the tests
then we may reject the uniformity of the sampler with high confidence.
However, if a sampler only fails one test then the sampler is likely
uniform and the failed test may be due to bad luck.

% We would like to follow with a discussion of the GOF test.
% he approximation to the $\chi^2$ distribution in the GOF
% test is known to break down if any expected number of occurrences
% is below five \citep{Yates1934ContingencyTI}.
% Thus the sample needs to be big enough to ensure that every expected occurrence
% count is at least five. Reducing the number of categories can (and in practice
% often does) reduce the size of the sample. Requiring a smaller
% sample size means fewer calls to a sampler which also boosts scalability.

\section{Threats to Validity}
\label{sec:threats}
As for any empirical study, there are some threats to consider. 

\subsubsection*{Internal Validity} As with any statistical test,
there is the possibility of wrongly rejecting or accepting $H_0$.
We mitigate the probability of falsely rejecting $H_0$
by choosing a low significance level $\alpha = 0.01$.
We believe that falsely accepting $H_0$
is not an issue considering our results (UniGen3 being the only
sampler that is deemed uniform by our tests and having
theoretical guarantees of uniformity).

\subsubsection*{External Validity} We cannot guarantee that our findings 
generalize to any formula and every sampler.
There are multiple reasons behind this. For example, a sampler could
be specifically engineered to overfit our battery of statistical tests without
being uniform which would require the development of new statistical tests.
Another example was presented in this paper with the synthetic formula benchmarks.
As shown in our results, the quality of the dataset is important
to have reliable results. As such, we cannot guarantee that there does not
exist a dataset $A$, such that a sampler would be uniform on the real-world
formulae used in this paper but not uniform on the dataset $A$.
To mitigate these threats, we selected multiple statistical tests
which test different properties of the sampler and a range of 
SAT formulae from multiple sources. 
The formulae encode different types of models: 
Electronic circuits, algorithmic problems, Linux kernels, Unix command line
tools or configuration tools, etc.
\citep{Chakraborty2014ug1,Chakraborty2015ug2,Dutra2018,Plazar2019,BURST,Halin2018}. 

\section{Conclusion}
\label{sec:conclusion}
To conclude,
we developed a series of statistical tests to test the practical uniformity
of uniform random samplers. By using these tests we have shown
that most sampler implementations do not produce samples with
uniform distribution thus demonstrating the need for more
systematic testing of uniform random samplers in general.
Systematic testing would allow sampler developers to catch bugs in their implementation
(if the sampler is not heuristic-based) before publication.
In addition, our battery of tests can be used to accelerate
the development of new samplers as theoretical proofs of uniformity
become less important.

Furthermore, we used synthetic formulae to demonstrate that the dataset
used to test the uniform random samplers is important. Thus we have
shown that before any testing is done, a dataset should be constructed
from a wide range of sources to achieve reliable results.
Sadly, these results also imply that exclusively using small formulae
to lower the computational cost of testing is not a good idea.
If a sampler is deemed non-uniform on a computationally easy dataset
then further testing is likely not necessary. If however the sampler
is deemed uniform, then further testing on more diverse and often more
computationally demanding datasets is likely necessary.

Finally, we would like to highlight that all our results
are available on our companion GitHub \citep{gitRepoCHI2}.
Adding samplers and statistical tests to the repository is possible
via pull requests thus creating a common playground
for future statistical tests and uniform random samplers.

% thanks
% \begin{acks}
\section*{Acknowledgements}
   This research was funded in whole, or in part, by the Luxembourg National Research Fund (FNR).
   Gilles Perrouin is a FNRS Research Associate. 
   Maxime Cordy and Olivier Zeyen are supported by FNR Luxembourg
   (grants
   
   \noindent
   C19/IS/13566661/BEEHIVE/Cordy and AFR Grant 17047437).
   %supported by the \grantsponsor{GS501100001809}{National Natural
   %Science Foundation of
   %China}{https://doi.org/10.13039/501100001809} under Grant
   %No.: ̃\grantnum{GS501100001809}{61273304}
   %and ̃\grantnum[http://www.nnsf.cn/youngscientists]{GS501100001809}{Young
   %Scientists' Support Program}.
% \end{acks}

%\section{Data Availability}
%\label{sec:data}
%\input{parts/9-Data_availability}

% \bibliographystyle{IEEEtran}
% \bibliographystyle{ACM-Reference-Format}
% \bibliographystyle{plainnat}
\bibliographystyle{dinat}
\balance
\bibliography{merged}

\end{document}